\documentclass[9pt,twocolumn,twoside]{osajnl}

\journal{ol} 

\usepackage{amsmath} 
\usepackage{siunitx}

\setboolean{shortarticle}{true}

\title{Miniaturized optical frequency standard for next-generation portable optical clocks}

\author[1,$\dagger$]{Vincent Maurice}
\author[1,$\dagger$,*]{Zachary L. Newman}
\author[2]{Susannah Dickerson}
\author[2]{Morgan Rivers}
\author[2]{James Hsiao}
\author[2]{Phillip Greene}
\author[2]{Mark Mescher}
\author[1]{John Kitching}
\author[1]{Matthew T. Hummon}
\author[2]{Cort Johnson}

\affil[1]{National Institute of Standards and Technology, Boulder, Colorado, 80305}
\affil[2]{The Charles Stark Draper Laboratory, Inc., Cambridge, Massachusetts 02139}

\affil[*]{Corresponding author: zachary.newman@nist.gov}
\affil[$\dagger$]{These authors contributed equally to this work.}

\begin{abstract}
Optical frequency standards, lasers stabilized to atomic or molecular transitions, are widely used in length metrology and laser ranging, provide a backbone for optical communications and lie at the heart of next-generation optical atomic clocks. Here we demonstrate a compact, low-power optical frequency standard based on the Doppler-free, two-photon transition in rubidium-87 at \SI[detect-weight=true]{778}{\nano\meter} implemented on a micro-optics breadboard. The optical standard achieves a fractional frequency stability of \num[detect-weight=true]{2.9e-12}/$\sqrt{\tau}$ for averaging times $\tau$ less than \SI[detect-weight=true]{e3}{\second}, has a volume of $\approx$\SI[detect-weight=true]{35}{\centi\meter^3} and operates on $\approx$\SI[detect-weight=true]{450}{\milli\watt} of electrical power. These results demonstrate a key step towards the development of compact optical clocks and the broad dissemination of SI-traceable wavelength references.
\end{abstract}

\setboolean{displaycopyright}{true}

\begin{document}

\maketitle

Since their development in the mid 1950s, atomic clocks have revolutionized  measurement science \cite{Evenson1972} and fundamental physics \cite{Hafele1962} and have enabled the development of critical technologies such as the global positioning system. The current generation of compact, low-power, atomic clocks rely on ground-state, microwave transitions in alkali atoms \cite{Knappe2004,Lutwak2011}. Although frequency standards based on lasers stabilized to optical transitions are widely recognized as being advantageous for precision timekeeping due to their high quality factors and relative insensitivity to environmental factors, field deployable optical clocks have not been pursued due to the added complexity of counting optical frequencies in a compact device. However, advances in the field of fiber-based \cite{Manurkar2018,Sinclair2014} and microresonator \cite{Pfeiffer2017,Briles2018,Raja2019,Stern2018,Suh2019} optical frequency combs 
bring a fully integrated optical clock within reach. Furthermore, compact optical standards could be used to realize SI-traceable references for frequency, wavelength, current and voltage outside the laboratory setting \cite{Kitching2016}{}.

Laboratory-scale optical atomic clocks based on laser-cooled, trapped atoms and ions achieve stabilities at a few parts in \num{e19} \cite{McGrew2019,Bothwell2019,Brewer2019}. These clocks derive their exceptional stability by probing doubly-forbidden, intercombination transitions in atoms with an ultra-stable laser. While such clock transitions offer extremely narrow linewidths ($\approx$\SI{1}{\milli\hertz} for strontium), accessing these states is experimentally challenging, requiring laser-cooling and trapping in an ultra-high vacuum chamber and prestabilization of the clock laser to a high-finesse optical cavity. 

Although they have significantly broader linewidths ($\approx$\num{e5} to \num{e7} \si{\hertz}), optical transitions in the alkali metals and simple molecules \cite{Hall1999,Schuldt2017}, such as the rubidium two-photon transition \cite{Grynberg1977}, are attractive candidates for a compact optical standard. These transitions can be addressed with commercially-available, narrow-linewidth sources and Doppler-free interrogation schemes can be employed that eliminate the need for laser cooling. Additionally, the fast relaxation times of excited atoms enables high-bandwidth feedback to the clock laser. In fact, a number of groups have demonstrated of vapor cell optical standards that achieve short-term stabilities competitive with (or outperforming) state-of-the-art, portable microwave clocks \cite{Esnault2011,Savory2018,AOSense}.

\begin{figure}[ht]
\centering
\includegraphics[width=\linewidth]{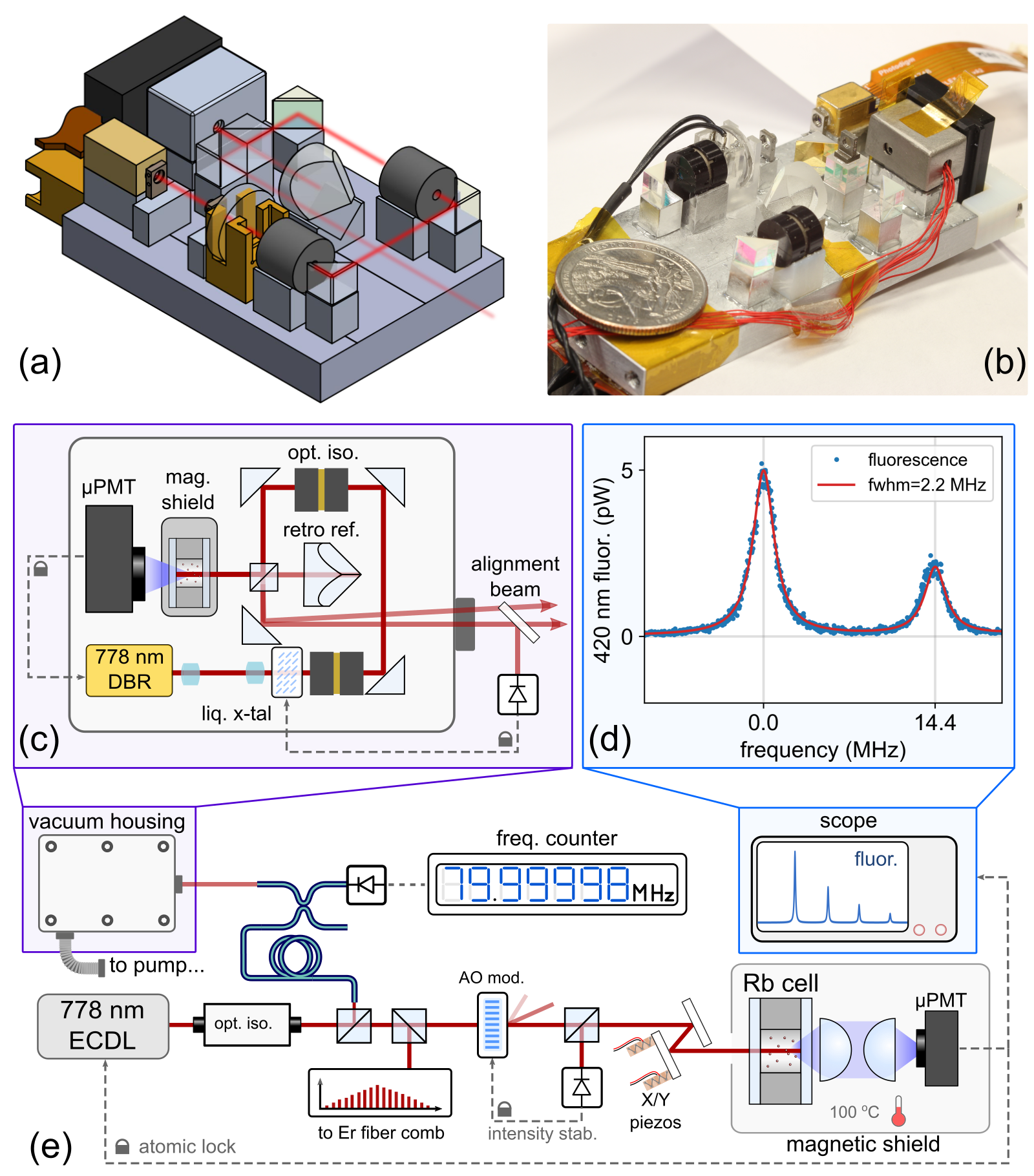}
\caption{Schematic (a) and image (b) of the miniature optical standard. The optical standard consists of a \SI{778}{\nano\meter} distributed-Bragg-reflector (DBR) laser, a series of miniature optical elements, an atom source (microfabricated vapor cell and shield) and a microfabricated photomultiplier tube (PMT). (c) Overhead schematic of the miniature physics package. The laser path is indicated by red arrows similar to (a). (d) Spectrum of the $5S_{1/2}\rightarrow5D_{5/2}$ two-photon transition in rubidium-87. (e) Optical layout of the measurement system.}
\label{fig:experiment}
\end{figure}

In this Letter, we introduce a miniaturized, low-power optical standard based on millimeter-scale optical components in which a semiconductor laser is stabilized to a microfabricated rubidium vapor cell. The $5S_{1/2}\mathrm{(F=2)}\rightarrow5D_{5/2}\mathrm{(F=4)}$, two-photon transition in rubidium-87 at \SI{778}{\nano\meter} serves as the frequency reference for our optical standard (described in detail elsewhere \cite{Hilico1998b,Poulin2002,Bernard2000,Martin2018}). Figure \ref{fig:experiment}(a) and (b) show the optical assembly that enables Doppler-free spectroscopy of the rubidium two-photon transition. The individual components that comprise the optical standard are off-the-shelf optics (with the exception of the cell and magnetic shield) that are held in custom-machined aluminum mounts, aligned using a micro-positioning system and set in place on an aluminum baseboard with a UV-curing epoxy.

Figure \ref{fig:experiment}(d) shows a spectrum (blue dots) of two of the hyperfine components (F = 4,3) of the clock transition as the laser frequency is swept across the resonance. We detect excitation of this transition via fluorescence at \SI{420}{nm} from the $5D\rightarrow 6P \rightarrow 5S$ decay, which is collected with a photomultiplier tube (PMT). A fit (red line) to the spectrum gives a linewidth of \SI{2.2}{\mega\hertz} (F=4). While this is substantially broader than the natural line width of \SI{330}{kHz}, it is not the limiting factor in the laser frequency stability.

The optical standard presented here employs an in-line geometry for probing and measuring the two-photon transition \cite{Newman2019}, in which the counter-propagating beams necessary for avoiding Doppler-broadening of the transition are generated by retro-reflecting the laser off a high reflectivity dielectric coating on the back of a planar, microfabricated cell (described below). Excitation of the transition is detected by collecting fluorescence directly behind the cell rather than from the side of the cell as is typically done in laboratory scale rubidium two-photon standards \cite{Bernard2000,Poulin2002,Martin2018}. This in-line detection allows us to place the PMT close to the cell, resulting in a compact geometry with good collection efficiency.

In order to orient the reader, we briefly describe the operation of the optical standard by detailing the optical path of the laser that is traced out by a red arrow in Fig. \ref{fig:experiment}(a) and (c). Light emitted from the DBR laser, which diverges sharply, is shaped with a telescope consisting of $f=$ \SI{1.45}{\milli\meter} and \SI{5}{\milli\meter} lenses. The lenses are arranged such that the laser is nominally focused at the back of the vapor cell with horizontal and vertical beam waists of $\approx$\SI{200}{\micro\meter} and $\approx$\SI{100}{\micro\meter}, respectively. For this beam waist, the light shift was measured to be $\approx$\SI{4.19}{\kilo\hertz/\milli\watt}.

Immediately following the telescope, the beam passes through a variable attenuator used for intensity stabilization comprised of a half waveplate, liquid crystal retarder, and a miniature optical isolator. The beam then passes through a pair of turning mirrors and through a second isolator to further reduce optical feedback to the laser. The isolators are positioned $\approx$\SI{2}{\centi\meter} apart so their individual magnetic fields do not interfere with each other and as far away from the vapor cell as possible (while maintaining a compact footprint). Following a third mirror, a 90/10, non-polarizing beam splitter directs ninety percent of the light into the vapor cell and passes the remaining ten percent onto a fourth mirror, which directs it away from the cell to serve as the 'output beam'. The output beam is used to characterize the laser frequency, monitor and stabilize the laser intensity and for alignment purposes.

Light entering the cell is reflected back towards the beam splitter by a high-reflectivity optical coating on the rear window of the cell. A fraction of this light passes through the beam splitter, is retro-reflected from a corner cube and ultimately follows the same path as the output beam (labeled the 'alignment beam' in Fig. \ref{fig:experiment}(c)). During construction, the forward and backward propagating beams that drive the two-photon transition are aligned by overlapping the output beam and the retro-reflected beam by adjusting the angle and position of the cell before it is epoxied in place. After assembly, the alignment beam to the corner cube is blocked.

\begin{figure}[htbp]
\centering
\includegraphics[width=\linewidth]{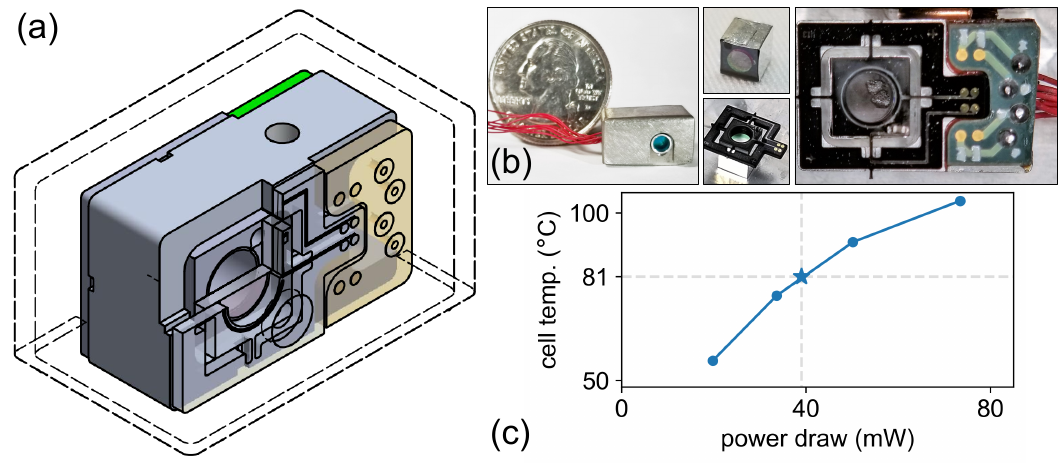}
\caption{(a) Schematic of atom source assembly (with dotted lines indicating the magnetic shield). The cut-away in the front left of the image reveals the silicon frame of the vapor cell and slices through the front tether and front window of the cell.  (b) Photograph of assembly and components including the magnetic shield (left), vapor cell and thermal tethers (middle top and bottom) and the full assembly (right). (c) Cell heater temperature versus power.}
\label{fig:cell}
\end{figure}

We measure the clock laser frequency by beating a portion of the output light against an external cavity diode laser (ECDL) locked to a second rubidium two-photon optical standard (Fig.\ref{fig:experiment}(e)). The frequency of the ECDL was shifted by \SI{80}{\mega\hertz} using an acousto-optical modulator and locked to the same hyperfine transition as the miniature optical standard. We detect the \SI{80}{\mega\hertz} beat using a standard silicon photodiode. The frequency stability of the ECDL was measured against a self-referenced, erbium-fiber frequency comb stabilized to a high finesse optical cavity and was found to be below \num{5e-13}/$\sqrt{\tau}$, where $\tau$ is the averaging time in seconds, and does not limit our measurements of the miniature optical standard.

Figure \ref{fig:cell} gives a detailed view of the atom source, which consists of a microfabricated vapor cell held on an aluminum mount. This mount is epoxied into the bottom half of a custom, single-layer magnetic shield. The shielding factor is $\approx$60 and is sufficient to prevent spectral broadening from the isolators. It is not sufficient, however, to prevent frequency shifts due to changes in the Earth’s magnetic field, which could be accomplished with a second shield surrounding the entire device. The cell is thermally isolated from the mount by a pair of laser-cut, polyimide tethers that suspend the cell in the middle of the aluminum frame, and it is heated by a pair of integrated platinum heater traces deposited on the tethers. The tethers and heaters are visible in Fig. \ref{fig:cell}(b). In order to avoid convective heat loss and reduce the total power consumption, the entire optical breadboard was housed in a small vacuum chamber (see Fig. \ref{fig:experiment}(d)). During measurements we operate the vacuum chamber near \num{e-7} \si{torr}, although pressures below \num{e-4} \si{torr} are sufficient to reduce thermal losses. The cell can reach a temperature of $\approx$\SI{100}{\hspace{2pt}\degree}C using only \SI{44}{\milli\watt} of electrical power when heated under vacuum. A more detailed power budget is given in the supplementary text.

\begin{figure}[htbp]
\centering
\includegraphics[width=\columnwidth]{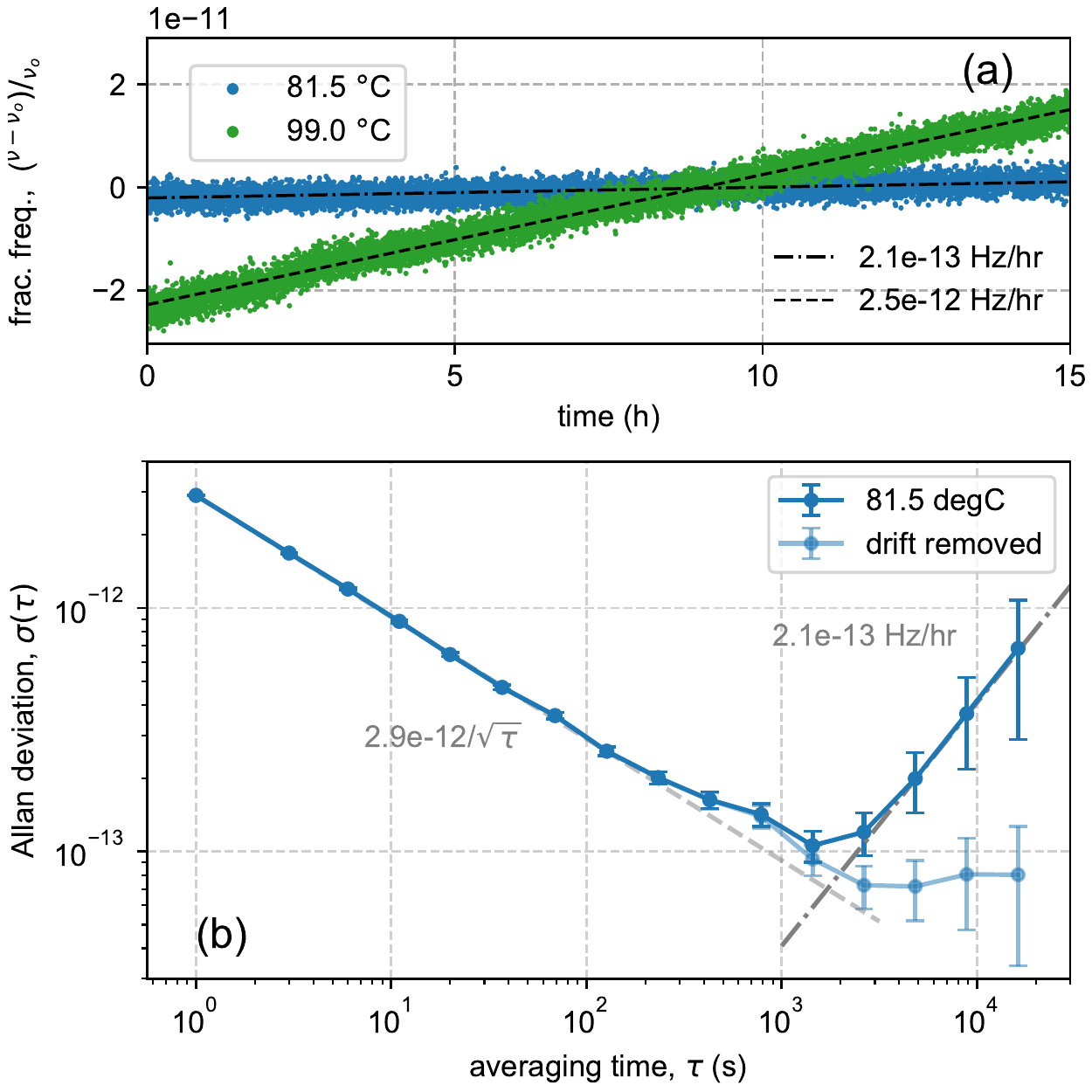}
\caption{Frequency stability of the miniature optical reference. (a) Time series of the optical reference taken with the cell temperature at \SI{81.5}{\hspace{1pt}\degree}C (blue) and \SI{99}{\hspace{1pt}\degree}C (green) during two different \SI{15}{hr} time windows.  Dashed lines indicate fitted drift rates. (b) Allan deviation of the optical reference at \SI{81.5}{\hspace{1pt}\degree}C.  Error bars represent 68\% confidence interval.}
\label{fig:adev}
\end{figure}

The vapor cell, shown in Fig. \ref{fig:cell}(b) is assembled by anodically bonding two pieces of glass to a silicon frame under vacuum \cite{Liew2004,Knappe2005b}. The front window is uncoated borosilicate glass and the back window is an aluminosilicate window with a high-reflectivity coating. Before bonding, the cell is filled in a nitrogen-purged glovebox by dipping a gold-coated copper wire in liquid rubidium metal and placing the wire inside the cell. A non-evaporable getter is also placed in the cell before bonding and is laser activated after bonding.

\begin{figure}[htbp]
\centering
\includegraphics[width=\linewidth]{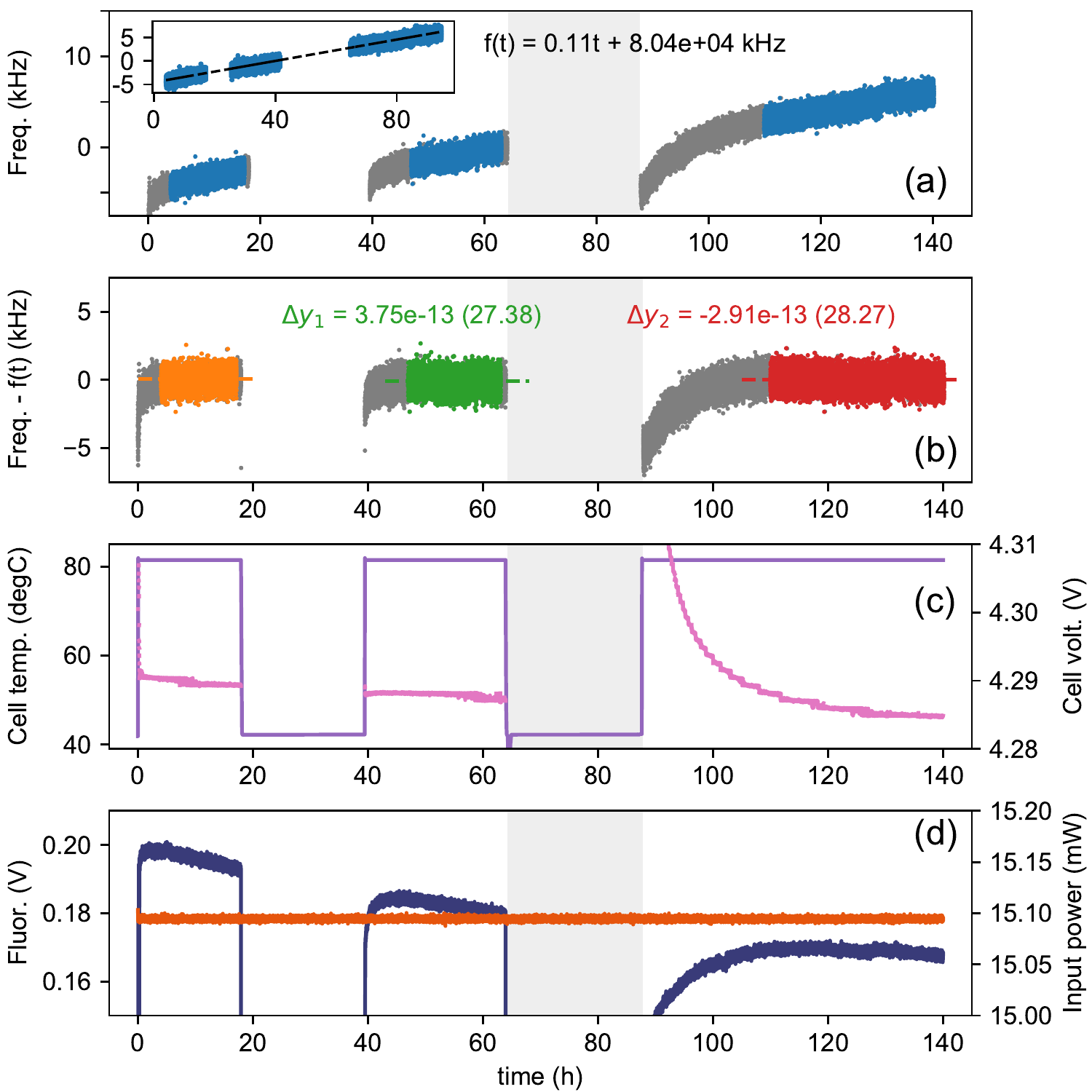}
\caption{Measurement of the clock laser frequency along with other system parameters over \SI{140}{\hour}. (a) Laser frequency during the measurement (grey) and the period where the frequency drifts linearly (blue). The inset shows the laser frequency only during the periods where $\mathrm{T_{cell}} =$ \SI{80}{\hspace{1pt}\degree}C and the frequency drifts linearly. (b) Laser frequency with the dirft rate calculated in the inset removed. (c) Cell temperautre and cell heater voltage. (d) Fluorescence level and in-loop input power. During the period between 65 and \SI{85}{\hour} (shaded light gray) the vacuum box was vented in order to test frequency shifts due to changes in the cell atmosphere.}
\label{fig:retrace}
\end{figure}

Figure \ref{fig:adev}(a) gives measurements of the laser frequency over a period of \SI{15}{\hour} for cell temperatures of \SI{81.5}{\hspace{1pt}\degree}C (blue) and \SI{99}{\hspace{1pt}\degree}C (green), which shows a clear temperature-dependent, linear drift. In both cases, the optical power seen by the atoms during the measurements was $\approx$\SI{15.1}{\milli\watt} (Fig. \ref{fig:retrace}(d), dark orange). Figure \ref{fig:adev}(b) gives the corresponding Allan deviation when the cell is operated at \SI{81.5}{\hspace{1pt}\degree}C. The short term stability of the optical standard is $\approx$\num{2.9e-12}/$\sqrt{\tau}$, which is limited by intermodulation noise \cite{Audoin1991} on the DBR laser.

Figure \ref{fig:retrace} shows a set of measurements designed to asses the temperatue sensitivity of the optical standard where we repeatedly dropped the cell temperature to \SI{40}{\hspace{1pt}\degree}C by reducing the cell heater current. During the measurements we monitored the laser frequency along with other system parameters including the in-loop laser power, cell heater voltage,  and the fluorescence level (Fig. \ref{fig:retrace}(d)). Figure \ref{fig:retrace}(a) shows the laser frequency over the course of the measurement where the breaks in the data correspond to periods where the cell temperature was \SI{40}{\hspace{1pt}\degree}C (we cannot operate the optical standard at low temperatures since the fluorescence signal used to lock the laser is too weak to engage the lock). The inset gives a plot of the laser frequency for $\mathrm{T_{cell}} =$ \SI{80}{\hspace{1pt}\degree}C as a function of the corrected time, i.e. only periods when the cell and baseplate temperatures have equilibrated and the frequency drifts linearly. We use this corrected time to determine a linear drift rate at \SI{80}{\hspace{1pt}\degree}C for the full measurement.  
Figure \ref{fig:retrace}(b) shows the laser frequency with the drift rate, $f(t)$, removed. The dotted lines show the average frequency (color coded) for each of the three periods where $\mathrm{T_{cell}} =$ \SI{80}{\hspace{1pt}\degree}C. The frequency difference between consecutive measurements ($\Delta\mathrm{y_x}$) is noted above the data. By removing the drift it becomes clear that there is no evolution in the laser frequency at low temperatures. In some sense, this can be considered a $\approx$\SI{24}{hr} retrace measurement as the entire system must thermally re-equilibrate and the laser must be relocked. In the absence of drift, a conservative estimate of the retrace is $\Delta\mathrm{y} = $\num{7.5e-13}. 
Figure \ref{fig:retrace}(d) gives a plot of the fluorescence (dark blue) which decays as the measurement progresses. We have considered a number of potential physical effects that could lead to a frequency shift and corresponding change in fluorescence including light shifts, the depletion of rubidium in the cell, evolution of background gases, migration of liquid-phase rubidium across the cell windows and the diffusion of helium through the cell windows. Ultimately we suspect the frequency drift is associated with misalignment of the interrogation beams at elevated temperatures as the epoxy securing the optical elements approaches its glass transition temperature. Changes in the baseplate temperature, measured indirectly via the cell heater voltage (Fig. \ref{fig:retrace}(c), pink) support this hypothesis. In addition, the misalignment shift is consistent with phenomena described elsewhere \cite{Martin2019} and with data presented in the supplementary text.

The advanced optical integration presented here demonstrates the feasibility of high-performance, compact, atomic clocks and wavelength references based on Doppler-free, optical transitions in warm atomic vapors. Even with the observed drift, the laser frequency stability is an order of magnitude better than commercially available microwave clocks of similar size at a fraction of the operating power \cite{Microsemi}. To minimize drift rates, devices employing ultra-stable \cite{Killow2013} or microfabricated \cite{Hummon2018} optical assemblies are needed. With further engineering, it is reasonable to expect that misalignment of the interrogation beams could be suppressed, resulting in significantly improved long-term frequency stabilities.

\section*{Funding Information}
Defense Advanced Research Projects Agency (DARPA); Atomic Clocks with Enhanced Stability (ACES).
\section*{Acknowledgments}
The authors would like to thank Jordan Stone and Liron Stern for comments on the manuscript. The views, opinions and/or findings expressed are those of the authors and should not be interpreted as representing the official views or policies of the Department of Defense or the U.S. Government. Any mention of commercial products within NIST web pages is for information only; it does not imply recommendation or endorsement by NIST. Distribution Statement "A" (Approved for Public Release, Distribution Unlimited).

\section*{Supplemental Documents}

See Supplementary Text for supporting materials. 


 

\bibliography{draperpp}

\end{document}